\documentclass[12pt]{article}
\usepackage{amsmath}
\usepackage{graphicx}
\usepackage{subfigure}
\usepackage{hyperref}

\begin{document}

\author{Ernst Trojan \textit{Moscow Institute of Physics and Technology} \and 
\textit{PO Box 3, Moscow, 125080, Russia}}
\title{Stability of hot tachyon gas}
\maketitle

\begin{abstract}
We consider a tachyon gas that obeys Maxwell-Boltzmann statistics. The sound
speed is always subluminal and it tends to the limiting minimum value $c_s=1/\sqrt{2}$ in non-relativistic gas (at low temperature), decreasing monotonously
with the growth of temperature and attaining ultra-relativistic limit $c_s=1/%
\sqrt{3}$ at high temperature. The hot tachyon gas always satisfies the
causality.
\end{abstract}

\section{Introduction}

The concept of tachyon fields plays significant role in the modern research,
where they often appear in the field theory, cosmology, theory of branes and
strings with various applications \cite{S2002,BBS2003,FKS02,D1,D2}.
Tachyons, are commonly known as field instabilities whose energy spectrum is 
\begin{equation}
\varepsilon _k=\sqrt{k^2-m^2}\qquad k>m  \label{t}
\end{equation}
where $m$ is the tachyon mass and relativistic units $c=\hbar =1$\ are used.

A system of many tachyons can be studied in the frames of statistical
mechanics \cite{M84,DHR89}, and thermodynamical functions of ideal tachyon
Fermi and Bose gases are calculated \cite{KRS07,KRS07b}.

We have recently considered a tachyon Fermi gas a zero temperature \cite
{TV2011c} and obtained a low-temperature expansion of its thermodynamical
functions at finite temperature \cite{T2011h} as well as thermodynamical
functions of fermionic thermal excitations \cite{TV2011d}. All this analysis
correspond to relatively low temperatures. At high temperature the Fermi
distribution is reduced to the Maxwell-Boltzmann distribution, and there is
no problem to consider and ideal Maxwell-Boltzmann gas of tachyons \cite
{DHR89}. However, the tachyon matter may occur unstable with respect to the
causality condition \cite{TV2011c}. So, the physical existence of hot
tachyonic matter is not evident until we check

If we consider a system of many tachyons as continuous medium, we need to
calculate the sound speed $c_s$ and establish the range of parameters when
the causality 
\begin{equation}
c_s\leq 1  \label{ca}
\end{equation}
is satisfied. Otherwise, the system will be unstable to hydrodynamical
perturbations and will not be able to exist in nature. In the present paper
we check the causality condition (\ref{ca}) for a Maxwell-Boltzmann gas of
tachyons.

\section{Thermodynamical functions}

Consider an ideal gas of free tachyons with the energy spectrum $\varepsilon
_p$ (\ref{t}) at finite temperature $T$. Let $\mu $ be the chemical
potential of this system. The particle number density $n$, energy density $E$
and pressure $P$ are determined by standard formulas \cite{TV2011c} 
\begin{equation}
n=\frac \gamma {2\pi ^2}\int\limits_m^\infty f_p\,p^2dp  \label{n}
\end{equation}
\begin{equation}
E=\frac \gamma {2\pi ^2}\int\limits_m^\infty f_p\varepsilon _pp^2dp
\label{e}
\end{equation}
\begin{equation}
P=\frac \gamma {6\pi ^2}\int\limits_m^\infty f_pp^3\frac{\partial
\varepsilon _p}{\partial p}d\varepsilon  \label{p}
\end{equation}
where $f_p$ is the distribution function. It is taken in the form of
Maxwell-Boltzmann distribution function 
\begin{equation}
f_p=\exp \left[ (\mu -\varepsilon _p)/T\right]  \label{f}
\end{equation}
for hot matter at high temperature. With dimensionless variables 
\begin{equation}
x=\frac{\varepsilon _k\,}T\qquad \beta =\frac mT  \label{x}
\end{equation}
the thermodynamical functions of tachyon matter (\ref{n})- (\ref{p}) are written in
the form: 
\begin{equation}
n=\frac{\gamma T^3}{2\pi ^2}\exp \left( \frac \mu T\right)
\int\limits_0^\infty \sqrt{x^2+\beta ^2}x\exp \left( -x\right) dx  \label{n1}
\end{equation}
\begin{equation}
E=\frac{\gamma T^4}{2\pi ^2}\exp \left( \frac \mu T\right)
\int\limits_0^\infty \sqrt{x^2+\beta ^2}x^2\exp \left( -x\right) dx
\label{e1}
\end{equation}
\begin{equation}
P=\frac{\gamma T^4}{6\pi ^2}\exp \left( \frac \mu T\right)
\int\limits_0^\infty \left( \sqrt{x^2+\beta ^2}\right) ^3\exp \left(
-x\right) dx  \label{p1}
\end{equation}
Integrating (\ref{p1}) by parts, we immediately find that 
\begin{equation}
P=nT  \label{p2}
\end{equation}
Hence 
\begin{equation}
E=nTJ\left( \frac mT\right)  \label{e2}
\end{equation}
where 
\begin{equation}
J\left( \beta \right) =\frac{\int\limits_0^\infty \sqrt{x^2+\beta ^2}x^2\exp
\left( -x\right) dx}{\int\limits_0^\infty \sqrt{x^2+\beta ^2}x\exp \left(
-x\right) dx}  \label{j}
\end{equation}

For relativistic Maxwell-Boltzmann gas of subluminal massive particle the
pressure and energy density are given by the same formulas (\ref{p2}) and (%
\ref{e2}) with the integral (\ref{j}) replaced by 
\begin{equation}
J\left( \beta \right) =\frac{\int\limits_0^\infty \sqrt{p^2+m^2}\exp \left( -%
\frac{\sqrt{p^2+m^2}}T\right) p^2dp}{\int\limits_0^\infty \exp \left( -\frac{%
\sqrt{p^2+m^2}}T\right) p^2dp}=\frac{\int\limits_\beta ^\infty \sqrt{%
x^2-\beta ^2}x^2\exp \left( -x\right) dx}{\int\limits_\beta ^\infty \sqrt{%
x^2-\beta ^2}x\exp \left( -x\right) dx}  \label{k}
\end{equation}
where $x=\varepsilon _p/T=\sqrt{p^2+m^2}/T$ and $\beta $ is defined
according to (\ref{x}). In the ultra-relativistic limit ($\beta \rightarrow 0$%
) both (\ref{j}) and (\ref{k}) tend to 
\begin{equation}
J\left( \beta \right) \rightarrow 3  \label{j3}
\end{equation}
that corresponds to ultra-relativistic equation of state $P=E/3$. Non-relativistic
approximation of (\ref{k}) correspond to $\beta =m/T \gg 1$, and it is no more than  
\begin{equation}
J\left( \beta \right) \rightarrow \beta +\frac 32=\frac mT+\frac 32
\label{j5}
\end{equation}
while for (\ref{j}) we have 
\begin{equation}
J\left( \beta \right) =2+\frac 3{2\beta ^2}=2+\frac 32\frac{T^2}{m^2}
\label{j2}
\end{equation}
Expressions (\ref{e2}) and (\ref{j5}) imply that 
\begin{equation}
E\rightarrow mn+\frac 32nT  \label{kk}
\end{equation}
for subluminal particles, while (\ref{e2}) and (\ref{j2}) imply 
\begin{equation}
E\rightarrow 2nT  \label{jj}
\end{equation}
for tachyons. The ratio 
\begin{equation}
w\left( \beta \right) =\frac PE=\frac 1{J\left( \beta \right) }  \label{w}
\end{equation}
at arbitrary $\beta $ is given in Fig.~\ref{ms1}.

\section{Sound speed}

The sound speed is defined as 
\begin{equation}
c_s^2=\left( \frac{\partial P\;}{\partial E}\right) _S  \label{cs}
\end{equation}
and we can calculated it for a tachyon gas if we apply the method of its
calculation for an ideal gas of subluminal particles \cite{LPPT83}.

We assume that the first law of thermodynamics is valid for tachyons: 
\begin{equation}
dU=-PdV+TdS  \label{sec}
\end{equation}
where 
\begin{equation}
U=EV  \label{u}
\end{equation}
is the internal energy, $V$ is the volume and $S$ is the entropy. The number
of particles$\ N=nV$ is conserved and does not depend on temperature, in
contrast to the non-conserved number of thermal excitations \cite{TV2011c}.
Then, we immediately define the specific heat per particle 
\begin{equation}
C_V=T\left( \frac{\partial \left( S/N\right) }{\partial T}\right) _V=\frac{%
\partial \left( U/N\right) }{\partial T}=\frac{\partial \left( E/n\right) }{%
\partial T}  \label{cv}
\end{equation}
that, in the light of (\ref{e2}) implies 
\begin{equation}
C_V=\frac{d\left( JT\right) }{dT}=J\left( \beta \right) -\beta J^{\prime
}\left( \beta \right)   \label{cv2}
\end{equation}
where 
\begin{equation}
J^{\prime }\left( \beta \right) =\frac{dJ}{d\beta }  \label{cv2b}
\end{equation}
and integral $J$ is defined by (\ref{k}) and by (\ref{j}) for massive
subluminal particles and for tachyons, respectively.

Equation (\ref{sec}) yields 
\begin{equation}
\left( \frac{\partial U}{\partial V}\right) _S=-P  \label{us}
\end{equation}
and 
\begin{equation}
\left( \frac{\partial U}{\partial V}\right) _T=T\left( \frac{\partial S}{%
\partial V}\right) _T-P  \label{ut}
\end{equation}
that in the light of (\ref{e2}) and (\ref{u}) implies 
\begin{equation}
\left( \frac{\partial S}{\partial V}\right) _T=\frac PT  \label{sv}
\end{equation}
Looking for adiabatic at $S={\rm const}$ in the form 
\begin{equation}
PV^\Gamma ={\rm const}  \label{ad}
\end{equation}
we find form (\ref{p2}) that 
\begin{equation}
\Gamma =1-\frac VT\left( \frac{\partial T}{\partial V}\right) _S  \label{g}
\end{equation}
Making transformation of derivatives \cite{LL54}, we have 
\begin{equation}
\left( \frac{\partial T}{\partial V}\right) _S=\frac{\partial \left(
T,S\right) }{\partial \left( V,S\right) }=\frac{\partial \left( T,S\right) }{%
\partial \left( T,V\right) }\frac{\partial \left( T,V\right) }{\partial
\left( V,S\right) }=-\left( \frac{\partial S}{\partial V}\right) _T\frac
T{C_V}  \label{ter}
\end{equation}
Substituting (\ref{sv}) and (\ref{ter}) in (\ref{g}) we obtain 
\begin{equation}
\Gamma =1+\frac 1{C_V}  \label{gg}
\end{equation}
In the light of (\ref{sec}), (\ref{u}) and (\ref{us}), equation (\ref{cs})
implies 
\begin{equation}
\left( \frac{\partial P\;}{\partial E}\right) _S=\left( \frac{\partial P\;}{%
\partial \left( U/V\right) }\right) _S=\frac{\;\Gamma }{1+U/\left( PV\right) 
}=\frac{\;\Gamma }{1+E/P}  \label{pes}
\end{equation}
Substituting (\ref{e2}), (\ref{p2}), (\ref{cv2}) and (\ref{gg}) in (\ref{pes}%
), we have 
\begin{equation}
c_s^2=\left\{ 1+\left[ \frac{d\left( TJ\right) }{dT}\right] ^{-1}\right\} 
\frac{1\;}{1+J}=\left( 1+\frac 1{J-\beta J^{\prime }}\right) \frac{1\;}{1+J}
\label{c}
\end{equation}
It is the sound speed in ideal gas with the Maxwell-Boltzmann distribution (%
\ref{f}).

Formula (\ref{c}) for non-relativistic subluminal particles (\ref{j5}) it
yields 
\begin{equation}
c_s^2=\frac 53\frac Tm  \label{c5}
\end{equation}
while for non-relativistic tachyons (\ref{j2}) the sound speed is 
\begin{equation}
c_s^2=\frac 12  \label{c11}
\end{equation}
and for ultra-relativistic material (\ref{j3}) the sound speed tends to the
same limit

\begin{equation}
c_s^2=\frac 13  \label{c1}
\end{equation}
The sound speed in tachyon gas at arbitrary temperature is shown in Fig.~\ref
{ms2}. It is clear that the causality condition (\ref{ca}) is always
satisfied.

\section{Conclusion}

\begin{figure}[tbp]
\caption{Specific heat per particle $C_V$ vs inverse temperature $\beta=m/T$
for tachyons (solid line) and subluminal particles (dashed line). }
\label{ms3}{\includegraphics[scale=0.8]{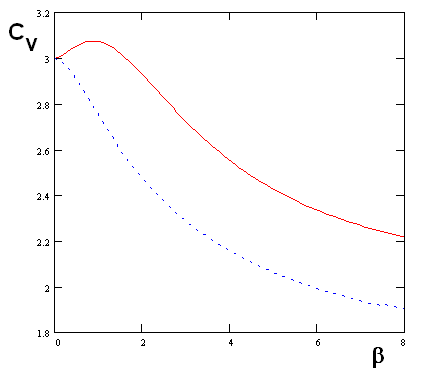}}
\par
For non-relativistic tachyons tachyons $C_V=2$ rather than $3/2$. 
\end{figure}

At high temperature an ideal gas of tachyons obey the Maxwell-Boltzmann
distribution (\ref{f}). Its pressure $P=nT$ (\ref{p2}) is given by the same
formula both for tachyons and ordinary particles, while the energy density
is given by formula $E=nTJ\left( T\right) $ (\ref{e2}), where integral $J$
for tachyons (\ref{j}) differs from that (\ref{k}) for subluminal particles.
The tachyonic specific heat is calculated
by formula (\ref{cv2}) and it is given in Fig.~\ref{ms3}. The tachyonic specific heat always exceeds the specific heat of ordinary gas, 
as well as the specific heat of tachyonic thermal excitations exceeds the specific heat of subluminal thermal excitations \cite{TV2011d}. 
However, the specific heat of hot tachyon gas reveals anomalous decrease with temperature only when $T>T_m\simeq 1.13m$ ($\beta <0.885$), while at $T<T_m$ the specific heat is growing with temperature. 

\begin{figure}[tbp]
\caption{Sound speed $c^2_s$ vs inverse temperature $\beta=m/T$ for tachyons
(solid line) and subluminal particles (dashed line). }
\label{ms2}{\includegraphics[scale=0.8]{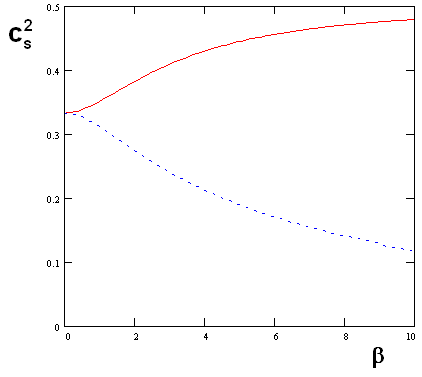}}
\par
For non-relativistic tachyons tachyons $c_s^2=1/2$. 
\end{figure}

The sound speed is calculated by formula (\ref{c}), it depends only on temperature $T$, and this dependence is given in Fig.~\ref{ms2}. 
The sound speed in hot tachyon gas (\ref{c}) is
always subluminal although its behavior differs form that in the ordinary
massive gas (Fig.~\ref{ms2}). 
It implies that the causality  (\ref{ca}) is automatically satisfied and the hot tachyon gas is
stable in the whole range of densities and temperatures, in contrast to the tachyon Fermi gas at zero tempearture \cite{TV2011c} and tachyonic
excitations \cite{TV2011d}.  
Therefore, the hot tachyon may have free surface where $P\rightarrow 0$ and may form compact
self-gravitating objects.

\begin{figure}[tbp]
\caption{Ration of pressure to energy density $P/E$ vs inverse temperature $%
\beta=m/T$ for tachyons (solid line) and subluminal particles (dashed line). 
}
\label{ms1}{\includegraphics[scale=0.8]{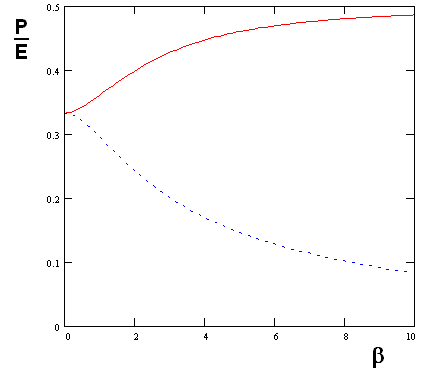}}
\par
For non-relativistic tachyons tachyons ($\beta \gg 1$) and $P=E/2$. 
\end{figure}

The pressure and density of hot tachyon always obey inequality $P<E/2$ (Fig.~%
\ref{ms1}), while a cold tachyon Fermi gas \cite{TV2011c} and a Fermi gas of
tachyonic thermal excitations \cite{TV2011d} can be 'hyperstiff' $P>E$.
However, the 'stiffness' of hot tachyon Fermi gas depends only on
temperature and does not depend on density, and the ration $P/E$ increases
when the temperature decreases (Fig.~\ref{ms1}), that bears resemblance with
the gas of tachyonic thermal excitations \cite{TV2011d} and never occurs in
the ordinary relativistic gas of subluminal particles. This strange behavior
may play important role in calculation of stellar models with tachyon
content.

Formula for the sound speed (\ref{c}) is universal and it can be developed for any
ideal gas whose pressure is given by proportionality $P\sim nT$. 
For exotic matter this problem deserves special consideration. 

The author is grateful to Erwin Schmidt for discussions.

\newpage

\end{document}